\documentclass[sigconf]{acmart}
\AtBeginDocument{%
  }

\setcopyright{acmlicensed}
\copyrightyear{2026}
\acmYear{2026}
\setcopyright{cc}
\setcctype{by}
\acmConference[ICPC '26]{34th IEEE/ACM International Conference on Program Comprehension}{April 12--13, 2026}{Rio de Janeiro, Brazil}
\acmBooktitle{34th IEEE/ACM International Conference on Program Comprehension (ICPC '26), April 12--13, 2026, Rio de Janeiro, Brazil}
\acmPrice{}
\acmDOI{10.1145/3794763.3794824}
\acmISBN{979-8-4007-2482-4/2026/04}

\usepackage{multirow,siunitx}
\usepackage{subcaption}

\newcommand{\best}[1]{\textbf{#1}} 
\newcommand{\negd}[1]{{\textcolor{red}{(\ensuremath{-}#1)}}}
\newcommand{\posd}[1]{{\textcolor{green}{(\ensuremath{+}#1)}}}
\newcommand{\zerod}{{\textcolor{gray}{(0.00)}}}

\begin{document}

\title{SkelDPO: A Skeleton-Guided Direct Preference Optimization Framework for Efficient Code Generation}

\author{Yu Yu}
\orcid{0009-0002-7096-2750}
\affiliation{%
  \department{School of Computer Science and Artificial Intelligence}
  \institution{Shandong Normal University}
  \city{Jinan}
  \country{China}
}
\email{2023028044@stu.sdnu.edu.cn}

\author{Chen Lyu}
\orcid{0000-0002-5044-1459}
\authornote{Corresponding author.}
\affiliation{%
  \department{School of Computer Science and Artificial Intelligence}
  \institution{Shandong Normal University}
  \city{Jinan}
  \country{China}
}
\email{lvchen@sdnu.edu.cn}

\renewcommand{\shortauthors}{Yu et al.}

\begin{abstract}
With the remarkable progress of Code Large Language Models (Code LLMs) in achieving semantic correctness, execution efficiency has become an increasingly important dimension for evaluating their practical utility. However, existing approaches typically treat full programs as a single optimization target during training, without explicitly modeling the structural factors that influence efficiency. As a result, although these models can generate semantically correct code, they fail to learn, at a fine-grained level, the underlying skeleton features that lead to efficient implementations.
To address this limitation, we propose SkelDPO (Skeleton-Guided Direct Preference Optimization), a skeleton-guided preference optimization framework that systematically enhances the efficiency of code generation. SkelDPO first identifies efficient and inefficient implementations from the code dataset and, through comparative analysis, locates their efficiency-prone and inefficiency-prone points, forming alignment signals between efficiency and inefficiency skeletons. During training, a joint code and skeleton preference loss is introduced, enabling the model to learn semantic correctness while reinforcing its understanding of efficiency-critical components in code.
We conduct comprehensive experiments on two efficiency-oriented benchmarks, Mercury and ENAMEL, using five representative backbone models. Results show that SkelDPO consistently surpasses existing methods: compared with SOTA method that relies solely on efficient and inefficient code preference optimization, it improves Pass@1, Beyond@1, and Effi@1 by 3–6\%, 3–7\%, and 2–5\%, with greater improvements observed on complex tasks. Further analysis demonstrates that SkelDPO exhibits strong robustness to hyperparameter settings, and higher proportions of skeleton-based samples lead to greater improvements. Overall, SkelDPO provides a new perspective on skeleton-level efficiency alignment, breaking the limitation of conventional preference optimization that relies solely on correctness or efficiency pairs, and paving the way for interpretable and transferable efficient code generation. All datasets and source code are publicly available at:
\url{https://github.com/YYYY-YuYu/SkelDPO}.
\end{abstract}




\maketitle

\section{Introduction}
In recent years, Large Language Models (LLMs) have made remarkable progress in code generation, enabling the direct transformation of natural language descriptions into executable programs~\cite{achiam2023gpt, openai2024gpt4o, claude4}. With the expansion of model scale and training data, Code LLMs such as Qwen2.5-Coder~\cite{hui2024qwen2.5coder}, StarCoder~\cite{li2023starcoder}, CodeLlama~\cite{roziere2023code}, and DeepSeek-Coder~\cite{guo2024deepseek} have achieved near-human performance on multiple benchmarks. However, the training objectives of most existing models remain primarily focused on functional correctness, i.e., generating programs that can pass test cases, while paying far less attention to runtime efficiency. In real-world software engineering scenarios, execution efficiency directly affects system responsiveness, resource utilization, and energy consumption. Consequently, efficient code generation has become an emerging and crucial direction for intelligent programming systems~\cite{cappendijk2024generating, peng2024large, shi2024efficient, feng2024llmeffichecker}.

Although substantial progress has been made in generating executable and correct programs, the efficiency aspect of code generation remains largely unsolved~\cite{niu2024evaluating, huang2024effilearner}. Current mainstream approaches—such as Supervised Fine-Tuning (SFT)~\cite{luo2023wizardcoder}, Reinforcement Learning with Human Feedback (RLHF)~\cite{wong2024aligning}, and Direct Preference Optimization (DPO)~\cite{zhang2024plum} typically treat correctness as the sole optimization target, neglecting the diverse and latent nature of efficiency signals. As a result, models often generate functionally correct but inefficient code, exhibiting redundant loop nesting, repeated computation, or frequent data-type conversions. These inefficiencies cannot be directly captured at the semantic level yet lead to substantial runtime differences. For example, in array manipulation or dynamic programming problems, models may adopt higher-complexity solutions or repeatedly invoke costly functions, resulting in multi-fold differences in execution time under equivalent logic.

To improve efficiency, several recent works have introduced performance-oriented training mechanisms. The EffiCoder~\cite{huangefficoder} series performs supervised fine-tuning on efficient programs to bias the model toward faster implementations, while CodeDPO~\cite{zhang2024codedpo} incorporates efficiency preference signals into the DPO framework to directly optimize the likelihood of generating efficient code. Although these methods yield moderate improvements in average runtime efficiency, they still suffer from three fundamental limitations. \textbf{First}, they treat the program as an indivisible optimization unit, ignoring the internal structures that critically determine runtime behavior. \textbf{Second}, their preference signals are coarse-grained, reflecting only overall speed differences and providing little guidance on the structural patterns underlying efficiency. \textbf{Third}, their optimization objectives remain single-dimensional, without establishing a stable trade-off between correctness and efficiency. Consequently, such models can generate efficient code on familiar tasks but exhibit unstable efficiency and limited generalization on complex or unseen problems.

Therefore, the central question we aim to address is: \textit{how can a model not only learn which code is faster, but also understand why certain implementations are more efficient?} Achieving this goal presents several challenges:
\textbf{First}, efficiency is a latent property that lacks direct supervision. Unlike syntax or semantics, runtime efficiency depends on local details—such as loop control, conditional evaluation, and data access patterns—that are not explicitly represented at the token level. Without fine-grained structural supervision, models can only mimic global performance results rather than learn the intrinsic mechanisms of efficient computation. To overcome this, we design a skeleton-level modeling mechanism that extracts shared substructures from efficient and inefficient implementations, forming abstract code skeletons that provide explicit structural supervision for local efficiency learning.
\textbf{Second}, functional correctness and structural efficiency are often in natural tension. Improving runtime performance typically requires restructuring the program—for example, introducing early loop termination, reducing temporary variables, or simplifying memory allocation—which may compromise semantic integrity or introduce corner-case errors. Hence, reinforcing performance signals alone can destabilize the balance between efficiency and correctness. We therefore adopt a multi-task joint optimization strategy that allows the model to simultaneously learn from correctness-oriented code generation and efficiency-oriented skeleton learning.
\textbf{Third}, existing preference optimization frameworks lack structural expressiveness. Traditional DPO compares efficient and inefficient programs only at the output level, ignoring transferable structural regularities that define efficient implementations. As a result, models can identify which code is faster but fail to generalize efficient patterns to unseen structures. To address this limitation, we introduce a skeleton-level preference signal into the DPO framework, achieving joint modeling of semantic and structural preferences and enabling the model to encode efficiency patterns beyond specific code instances.

Building upon these insights, we propose \textbf{Skel}eton-guided \textbf{D}irect \textbf{P}reference \textbf{O}ptimization (SkelDPO), a novel framework that extends conventional DPO by incorporating skeleton-guided structural preference modeling. SkelDPO expands efficiency alignment into two complementary levels: the code level, which captures explicit runtime differences between executable implementations, and the skeleton level, which models abstract regularities shared among efficient implementations. Through joint optimization across both levels, the model maintains semantic correctness while refining internal structure distributions, thereby achieving a unified balance between efficiency and interpretability. The core idea of SkelDPO is to enable the model to learn not only the results but also the forms of efficient implementations. Instead of optimizing only for overall performance, SkelDPO decomposes the efficiency signal into semantic and structural components, allowing the model to increase the probability of correct solutions while gradually biasing toward efficient skeletons. The framework requires no additional reward model or reinforcement learning pipeline, ensuring stable training, strong transferability, and easy integration with existing Code LLMs.

In addition, SkelDPO incorporates a systematic data construction pipeline. We extract efficient and inefficient implementations from the APPS dataset and automatically generate paired code and skeleton samples through comparative analysis. Skeletons are obtained by shared-token extraction and structural alignment, accurately highlighting the divergent efficiency regions between implementations. This structured data construction not only stabilizes preference supervision but also provides cross-code representations of efficient coding patterns.

Experimental results demonstrate that SkelDPO achieves significant and consistent improvements across multiple models and benchmarks. On Mercury and ENAMEL, SkelDPO improves Pass@1 by 4–8\% and efficiency metrics (Beyond@1, eff@1) by 5–10\%, while achieving stronger stability and generalization on complex tasks such as the Interview and Hard categories. 

The main contributions of this paper are as follows:
\begin{itemize}
\item \textbf{Skeleton-guided preference optimization framework (SkelDPO).}  
We propose SkelDPO, which introduces a skeleton preference signal into the traditional DPO framework. 
By jointly optimizing code and skeleton preferences, the model retains functional correctness while gaining structural inductive capability, thereby significantly improving code generation efficiency.

\item \textbf{Fine-grained dataset of efficient and inefficient code with aligned skeletons.}  
We construct a fine-grained dataset by screening and aligning efficient and inefficient implementations across multiple tasks. 
This dataset provides directly usable sample pairs for skeleton preference optimization and establishes a new foundation for future efficiency-oriented research.

\item \textbf{Extensive experimental validation across multiple models and benchmarks.}  
Comprehensive experiments on Mercury and ENAMEL demonstrate that SkelDPO significantly outperforms existing approaches (such as EffiCoder and CodeDPO) and exhibits superior generalization on complex tasks.
\end{itemize}
\section{Motivation}

\subsection{Existing Problems and Observations}

With the rapid development of Large Language Models (LLMs) for code generation, models have made remarkable progress in functional correctness. Current Code LLMs, such as StarCoder~\cite{lozhkov2024starcoder}, CodeLlama~\cite{roziere2023code}, Qwen2.5-Coder~\cite{hui2024qwen2.5coder}, and DeepSeek-Coder~\cite{guo2024deepseek}, can stably produce programs that pass unit tests, greatly improving developer productivity. However, correctness does not imply efficiency. In practical applications, runtime efficiency directly affects system performance and resource costs, yet existing models still exhibit significant limitations in this dimension.

\begin{figure}[h]
    \centering
    \includegraphics[width=1\columnwidth]{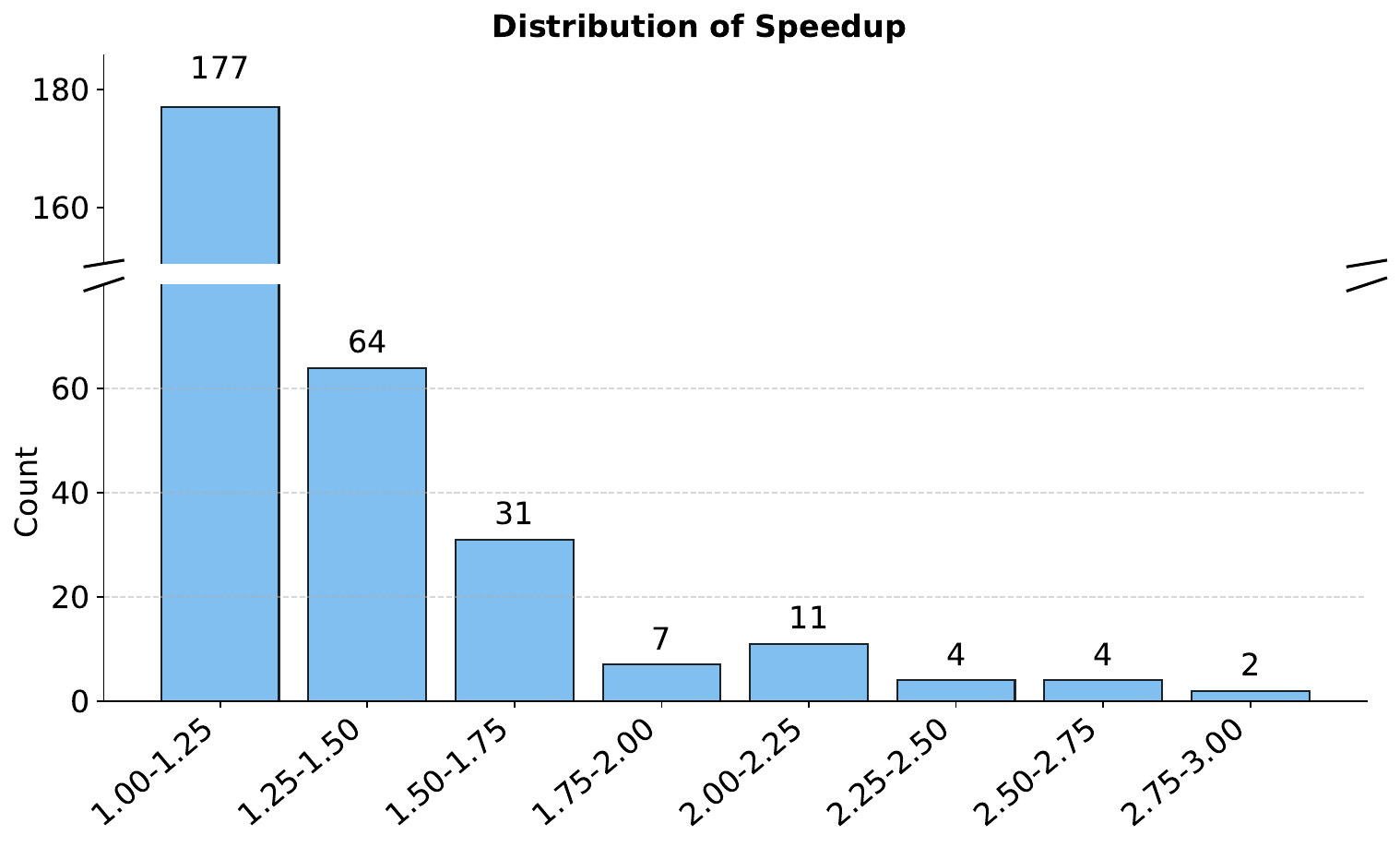}
    \caption{Distribution of runtime speedup ratios among correct program implementations on the APPS dataset. Each bar shows the number of solutions falling within a specific speedup range (relative to the slowest correct code).}
    \label{fig:speedup-dist}
\end{figure}

To verify this issue, we conducted a preliminary experiment on the APPS~\cite{hendrycks2021measuring} dataset using Qwen2.5-Coder-1.5B-Instruct. For a randomly selected problem, we generated 300 correct program implementations and executed them under identical input conditions. Despite producing the same outputs, their runtime efficiency varied substantially—the fastest and slowest implementations differed by approximately \textbf{1–3$\times$} in execution time (Figure~\ref{fig:speedup-dist}). The majority of programs achieve a speedup between 1.0$\times$ and 1.5$\times$, while a small fraction reaches up to 3$\times$, confirming that correct solutions vary substantially in execution efficiency.

For instance, in an array-deduplication task, all programs returned correct results, but the version using a built-in set() required only 3 ms, while the nested-loop version that manually checked duplicates took 27 ms. Such disparities are not isolated; they occur in most tasks. This observation shows that existing models can “write correct code” but have not learned to “write efficient structures.”

\begin{figure}[h]
    \centering
    \includegraphics[width=0.9\linewidth]{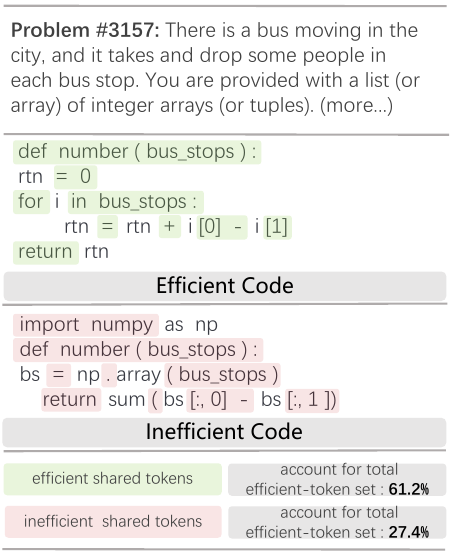}
    \caption{Structural similarity comparison between efficient and inefficient implementations.}
    \label{fig:motivation}
\end{figure}

\subsection{Empirical Analysis and Motivating Experiments}

To better understand the relationship between code efficiency and structural design, we selected 100 tasks from the APPS dataset, each with ten executable correct implementations. We measured the average runtime of each implementation, ranked them by performance, and defined the top 5 as efficient code and the bottom 5 as inefficient code.
We then compared their structural similarity by computing the shared-token ratio (i.e., structural overlap) between implementations. 
As shown in Figure~\ref{fig:motivation}, the shared tokens among efficient implementations account for \textbf{61.2\%} of the total efficient-token set, whereas those shared with inefficient implementations account for only \textbf{27.4\%}.
This striking contrast indicates that efficient programs share highly stable and consistent structures, whereas inefficient ones are far more diverse.

To further examine whether such structural skeletons can transfer efficiency, we extracted the efficient and inefficient skeletons from each task and used them as explicit prompts to guide code generation with Qwen2.5-Coder-1.5B-Instruct. Specifically, each skeleton was appended to the original problem description and provided to the model as a prompt to generate new solutions. 
We then re-executed the generated programs and measured their average runtime. Results show that incorporating efficient skeletons led to a \textbf{1.18$\times$ speedup} compared to the baseline
generation, while using inefficient skeletons caused the runtime to \textbf{decrease by 0.92$\times$}. This experiment reveals a crucial insight: the structural features of efficient code are transferable, and the skeleton itself encapsulates reusable efficiency information.

However, existing Supervised Fine-Tuning(SFT) and Direct Preference Optimization(DPO) frameworks generally treat the entire program as a single optimization unit, lacking explicit modeling of the efficiency or inefficiency of specific code segments.
In SFT, the training objective mainly focuses on the functional correctness or overall runtime efficiency of the entire code, while neglecting the critical role that certain tokens play in performance differences. DPO is similar: although it introduces preference information between incorrect and inefficient code, the optimization remains confined to the program level, without explicitly identifying or abstracting the key code skeletons that drive efficiency improvement. This characteristic—"learning overall code efficiency without learning key efficient tokens"—makes it difficult for the model to understand the internal mechanisms of code efficiency.

These issues together reveal a core conclusion: improving runtime efficiency requires not only distinguishing efficient and inefficient programs at the overall level but also explicitly modeling shared high-efficiency tokens during training. Only by directly exposing the model to the structural abstractions of efficient code and allowing it to learn their internal patterns in the optimization process can the model truly learn to generate efficient code.

\label{sec:skeldpo}

\begin{figure*}[t]
    \centering
    \includegraphics[width=\textwidth]{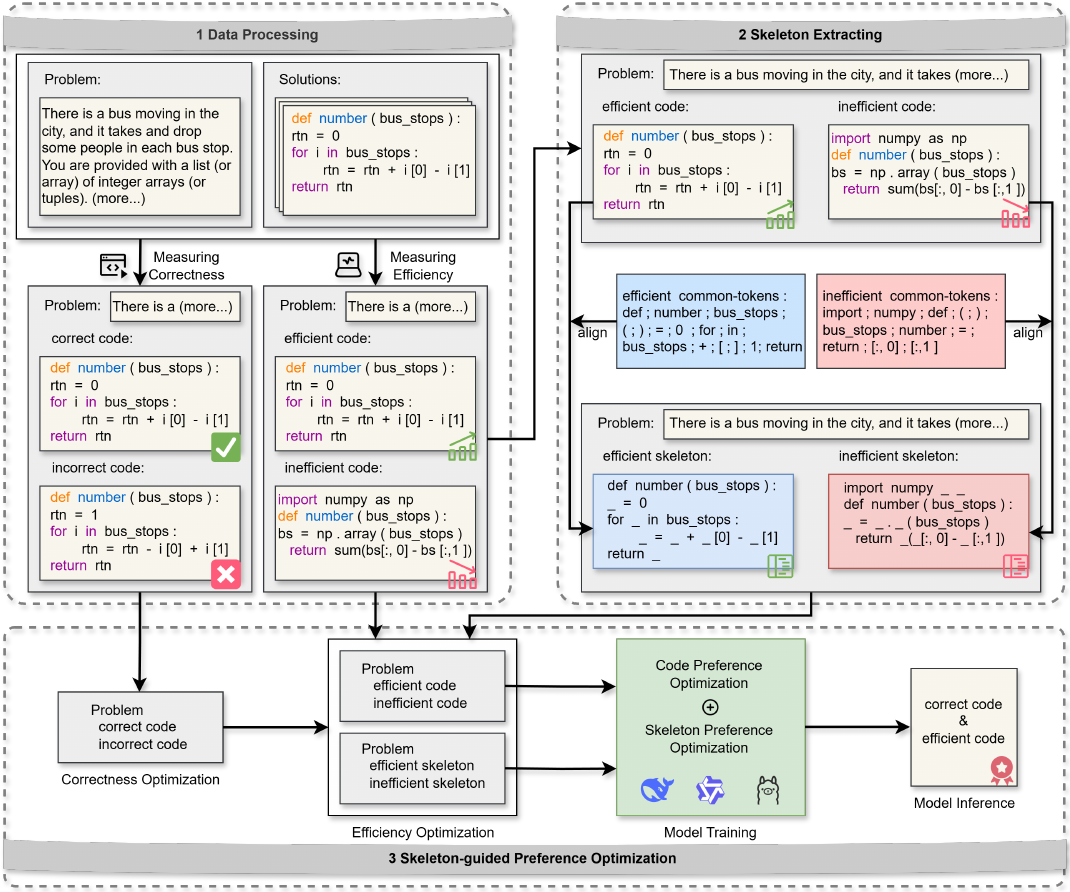}
    \caption{
        Overall framework of the Skeleton-guided Preference Optimization (SkelDPO) approach.
        The framework includes three stages:
        (1) \textbf{Data Processing}, where the execution results are used to identify correct and efficient code;
        (2) \textbf{Skeleton Extracting}, which derives efficient and inefficient code skeletons;
        and (3) \textbf{Skeleton-guided Preference Optimization}, which combines code-level and skeleton-level signals to fine-tune the LLM toward generating correct and efficient code.
    }
    \label{fig:framework}
\end{figure*}

\section{Focused Preference Optimization via SkelDPO}

We propose SkelDPO (Skeleton-guided Direct Preference Optimization), a framework that performs Focused Preference Optimization to steer the model’s attention during training toward code skeleton thereby improving both the quality and efficiency of generated programs. Unlike conventional DPO that contrasts preferences over whole programs, SkelDPO applies fine-grained optimization only to performance-critical fragments, enabling the model to better capture and reinforce efficient coding patterns. As illustrated in Figure~\ref{fig:framework}, SkelDPO comprises three core stages: data processing, skeleton extracting and skeleton-guided preference optimization




\subsection{Data Processing}

Let $x_i$ denote the problem description, $y_i$ a candidate code solution, and $t_i$ the associated unit tests. For each problem, we execute all candidates against $t_i$ to obtain functional correctness and runtime measurements.

\paragraph{(1) Measuring correctness and runtime.}
Each candidate $y_i$ undergoes an automated evaluation along two dimensions:
(i) \textbf{Correctness}: whether $y_i$ passes all tests in $t_i$;
(ii) \textbf{Runtime}: stable execution-time measurements via repeated runs.

Procedure.
(1) \textbf{Repeated execution and sampling.} Each $y_i$ is executed $N{=}50$ times, recording outcome and time (nanosecond precision). We compute the mean $\mu$ and standard deviation $\sigma$, and the coefficient of variation $\mathrm{CoV}=\sigma/\mu$ to assess stability.  
(2) \textbf{Stability criterion.} If $\mathrm{CoV}<0.01$, the measurement is deemed stable in spirit to Code-Optimise~\cite{gee2024code}.  
(3) \textbf{Outliers and exclusion.} If stability is not achieved after retries, the solution is marked unreliable and excluded.  
(4) \textbf{Cross-process repeats and final aggregation.} To further reduce interference, the measurement pipeline is repeated in $5$ independent processes; the average across the $10$ rounds is used as the final runtime metric.

\paragraph{(2) Selecting efficient/inefficient samples.}
For each problem $x_i$, we sort all correct solutions by runtime and retain at least $10$ correct ones. The fastest $5$ are labeled efficient solutions and the slowest $5$ inefficient solutions, yielding paired contrasts with a clear performance gradient.

\paragraph{(3) Cleaning and filtering.}
We discard problem instances with fewer than $10$ correct or fewer than $5$ slow (inefficient) solutions, avoiding bias and sparsity. 

\subsection{Skeleton Extracting}

Given the efficient/inefficient sets, we perform fine-grained analysis to identify code regions with significant impact on runtime, i.e., code skeletons. For each problem, we choose the fastest, fully correct implementation as the ground-truth implementation and define the tests it passes as the ground-truth test set. Candidates that pass all ground-truth tests are partitioned into efficient and inefficient sets.

Within each problem, we first extract the common token sets from efficient and inefficient implementations, and then construct their corresponding skeleton representations.  
Specifically, we tokenize all efficient solutions and compute their intersection to obtain the \emph{efficient common-token set} $\mathcal{T}^{+}$; likewise, we obtain the \emph{inefficient common-token set} $\mathcal{T}^{-}$:
\[
\mathcal{T}^{+} = \bigcap_{y \in \mathcal{Y}_{\text{eff}}} \mathrm{Token}(y),
\qquad
\mathcal{T}^{-} = \bigcap_{y \in \mathcal{Y}_{\text{ineff}}} \mathrm{Token}(y),
\]
where $\mathrm{Token}(\cdot)$ denotes the tokenization function.  
Given $\mathcal{T}^{+}$ and $\mathcal{T}^{-}$, we align them back to each original code sequence to construct the corresponding \emph{efficient skeleton} $S^{+}$ and \emph{inefficient skeleton} $S^{-}$.  
For each token in the original code, if it belongs to $\mathcal{T}^{+}$ (or $\mathcal{T}^{-}$), we keep it; otherwise, we replace it with a special placeholder token \texttt{<MASK>}.  
This masking process preserves the shared structural components while abstracting away implementation-specific details, yielding high-level skeleton representations that capture efficiency-related patterns.

$S^{+}$ encodes structural regularities commonly present in efficient implementations, whereas $S^{-}$ highlights redundant or inefficient structures frequently observed in slower ones.  
By contrasting $S^{+}$ and $S^{-}$, we can localize efficiency-critical regions—such as loop control, data-structure manipulation, and redundant computations—that strongly influence runtime behavior.  
After filtering and cleaning, we obtain $2{,}880$ training problems, $28{,}800$ correct/incorrect code pairs (10 correct solutions per problem), $2{,}880$ efficient/inefficient code pairs and $2{,}880$ efficient/inefficient skeleton pairs for downstream training.

\subsection{Skeleton-guided Preference Optimization}

With paired code samples and their skeletons, we develop SkelDPO, which jointly learns efficiency preferences at the code and skeleton levels, reinforcing efficient patterns during generation without external reward models. SkelDPO builds on Direct Preference Optimization (DPO). Standard DPO optimizes over pairs $(y^{+},y^{-})$ by encouraging the model to increase the probability of $y^{+}$ relative to $y^{-}$ for the same input $x$ via a contrastive objective. A common instantiation is:

\begin{equation}
\label{eq:dpo-split}
\begin{split}
\mathcal{L}_{\mathrm{DPO}}(\theta)
= -\,\mathbb{E}_{(x,y^{+},y^{-})}\bigg[
  \log \sigma\!\Big(\beta\big(
   \underbrace{\log \pi_{\theta}(y^{+}\!\mid x)-\log \pi_{\theta}(y^{-}\!\mid x)}_{\text{model logit diff}} \\
   -\,\underbrace{\log \pi_{\mathrm{ref}}(y^{+}\!\mid x)+\log \pi_{\mathrm{ref}}(y^{-}\!\mid x)}_{\text{reference adjustment}}
  \big)\Big)\bigg].
\end{split}
\end{equation}

where $\pi_{\theta}$ is the current model, $\pi_{\mathrm{ref}}$ is a fixed reference model, $\beta{>}0$ is a temperature, and $\sigma(\cdot)$ is the sigmoid.

\paragraph{Code-level preference loss $\mathcal{L}_{\text{code}}$.}
We apply DPO to full, executable programs using pairs $(y^{+},y^{-})$ that differ in efficiency while holding correctness comparable:
\begin{equation}
\label{eq:lcode}
\mathcal{L}_{\text{code}}(\theta)
= \mathcal{L}_{\text{DPO}}\!\big(\theta; (x, y^{+}, y^{-})),
\end{equation}

This encourages the model, for a given $x$, to prefer efficient implementations among semantically correct alternatives.
The prompt used for efficient code generation is shown below:
\begin{verbatim}
[GEN_CODE]
INSTRUCTION: Please generate an efficient and correct 
program that solves the given problem.
QUESTION: <problem description>
ANSWER:
\end{verbatim}

\paragraph{Skeleton-level preference loss $\mathcal{L}_{\text{skeleton}}$.}
To endow the model with structural efficiency priors, we introduce a skeleton-level DPO objective over pairs $(S^{+},S^{-})$:
\begin{equation}
\label{eq:lskeleton}
\mathcal{L}_{\text{skeleton}}(\theta)
= \mathcal{L}_{\text{DPO}}\!\big(\theta; (x, S^{+}, S^{-})),
\end{equation}

Compared with full-code pairs, this objective is more abstract and structurally aggregated, strengthening the model’s tendency to realize efficient skeletons during generation.
The prompt used for code skeleton generation is shown below:
\begin{verbatim}
[GEN_CODESKELETON]
INSTRUCTION: Please generate the efficient code skeleton 
that reflects high-performance patterns of implementation.
QUESTION: <problem description>
Let's think by code skeleton:
\end{verbatim}

\paragraph{Joint objective.}
For each problem, we construct two kinds of preference pairs: $(y^{+},y^{-})$ at the code level and $(S^{+},S^{-})$ at the skeleton level. SkelDPO jointly models both signals with
\begin{equation}
\label{eq:joint}
\mathcal{L}_{\text{SkelDPO}}(\theta)
= \alpha \cdot \mathcal{L}_{\text{code}}(\theta)
+ (1-\alpha) \cdot \mathcal{L}_{\text{skeleton}}(\theta), \alpha \in [0,1].
\end{equation}
Here, $\mathcal{L}_{\text{DPO}}(\cdot)$ is the standard DPO loss over the corresponding pairs, and $\alpha$ balances the code- and skeleton-level contributions. 

\section{Experimental Setup}
\label{sec:exp_setup}

\subsection{Research Questions}

To systematically evaluate the effectiveness of \textbf{SkelDPO} in efficiency-oriented code generation, we design five research questions (RQs):

\textbf{RQ1 (Overall Performance).}  
\emph{Can SkelDPO outperform existing efficiency-oriented baselines in both correctness and efficiency?}  

\textbf{RQ2 (Cross-Reference Model Generalization).}  
\emph{Can SkelDPO remain effective when applied to different reference models with varying pretraining quality and task alignment?}  

\textbf{RQ3 (Parameter Sensitivity).}  
\emph{How does the relative weighting between code-level and skeleton-level preference optimization losses influence the final model performance?}  

\textbf{RQ4 (Skeleton Data Scaling).}  
\emph{How does scaling the proportion of skeleton-guided data impact performance?}  

\textbf{RQ5 (Qualitative Analysis).}  
\emph{Compared with CodeDPO, how does SkelDPO affect the structure, readability, and efficiency of real generated programs?}  

\subsection{Datasets and Metrics}

\textbf{Datasets.}
Mercury~\cite{du2024mercury} (from LeetCode) contains \textbf{256} evaluation tasks with effectively unbounded test cases (\(+\infty\)), targeting algorithmic Python problems.  
ENAMEL~\cite{qiu2024efficient} (from HumanEval) includes \textbf{142} tasks with about \textbf{20} tests per task, focusing on general-purpose Python coding.

\textbf{Metrics.}
Pass@1 measures single-sample functional correctness.  
Beyond@1 measures the relative runtime efficiency percentile of a correctly executed single generated program among all reference implementations for the same task.  
eff@1 evaluates the normalized efficiency score of a single correct generation relative to human expert implementations. It measures runtime performance across standardized test cases and statistically adjusts for right-censored timeouts.  

\subsection{Base Models and Baselines}

\textbf{Models.}
To thoroughly assess SkelDPO’s universality and effectiveness, we evaluate it on five representative code LLMs:
Qwen2.5-Coder-1.5B-Instruct~\cite{hui2024qwen2.5coder}, StarCoder2-3B~\cite{lozhkov2024starcoder}, DeepSeek-Coder-6.7B-Instruct~\cite{guo2024deepseek}, CodeLlama-7B-Python-hf~\cite{roziere2023code} and Qwen2.5-Coder-7B-Instruct~\cite{hui2024qwen2.5coder}

\textbf{Baselines.}
We compare SkelDPO with four representative baselines:

\textbf{Instruct (Base Model)}\cite{ye2025llm4effi}: an instruction-tuned code model used with standard prompting.

\textbf{LLM4Effi (Prompt-based)}\cite{ye2025llm4effi}: a prompting-based framework that prioritizes efficiency via logic-domain algorithm exploration and code-domain implementation optimization, then refines correctness with checked synthetic 

\textbf{EffiCoder (SFT-based)}\cite{huangefficoder}: a supervised fine-tuning method trained on efficient solutions curated by profiling execution time improving both correctness and runtime efficiency.

\textbf{CodeDPO (DPO-based)}\cite{zhang2024codedpo}: a direct preference optimization framework that builds preference pairs and aligns models to prefer code that is both more correct and more efficient at the whole-program.

\subsection{Training and Inference Settings}

\textbf{Training.}
SkelDPO is trained using the LoRA (Low-Rank Adaptation) fine-tuning scheme to reduce parameter overhead and stabilize training, which is widely used in prior work \cite{du2024mercury, paul2025obscuracoder}.  
Batch size is set to 64, learning rate to $1\times10^{-5}$, with abeta of 0.1 and training of 800 epochs.  
The weights of code and skeleton preference losses are $\alpha{=}0.8$ and $(1-\alpha){=}0.2$.  
All experiments are conducted on a cluster equipped with 4 RTX 5880-48GB GPUs.

\textbf{Inference.}
During inference, we adopt a unified decoding setup with temperature=0.8, top\_p=0.95, and maximum generation length of 1024 tokens with zero-shot.  
All outputs are executed in an isolated sandbox to measure runtime and correctness.  
All experiments share identical hardware and sandbox configurations to ensure reproducibility and fair comparison.
\section{Experiments and Results}
\label{sec:exp_results}

\subsection{RQ1: Overall Performance}

We first compare SkelDPO with mainstream efficiency-oriented code generation baselines, including Instruct, LLM4Effi, EffiCoder, and CodeDPO, on the Mercury and ENAMEL benchmarks. All models are trained and evaluated under identical settings using the same dataset metrics (Pass@1, Beyond@1, and eff@1), as shown in Table~\ref{tab:main_results}.

\begin{table}[h]
\footnotesize
\caption{Overall results on \textbf{Mercury} and \textbf{ENAMEL}. 
We report Pass@1, Beyond@1, and eff@1 (\%). Best numbers per backbone are in \textbf{bold}.}
\label{tab:main_results}
\centering
\resizebox{\linewidth}{!}{
\begin{tabular}{l l c c c c}
\toprule
\multicolumn{2}{c}{\textbf{Model / Method}} &
\multicolumn{2}{c}{\textbf{Mercury}} &
\multicolumn{2}{c}{\textbf{ENAMEL}} \\
\cmidrule(lr){3-4}\cmidrule(lr){5-6}
 &  & \textbf{Pass@1} & \textbf{Beyond@1} & \textbf{Pass@1} & \textbf{eff@1} \\
\midrule
\multirow{5}{*}{Qwen2.5–1.5B}
  & Instruct   & 26.17 & 18.07 & 21.13 & 13.73 \\
  & LLM4Effi   & 28.91 & 20.88 & 23.24 & 16.12 \\
  & EffiCoder  & 37.89 & 25.50 & 29.57 & 20.69 \\
  & CodeDPO    & 38.28 & 28.19 & 36.62 & 20.68 \\
  & \textbf{SkelDPO} & \textbf{41.41} & \textbf{29.60} & \textbf{39.44} & \textbf{23.15} \\
\midrule
\multirow{5}{*}{StarCoder2-3B}
  & Instruct   & 11.72 &  7.35 &  7.75 &  3.13 \\
  & LLM4Effi   & 13.67 &  8.22 &  9.15 &  6.69 \\
  & EffiCoder  & 42.58 & 28.91 & 12.68 &  6.70 \\
  & CodeDPO    & 43.36 & 29.14 & 16.20 & 11.61 \\
  & \textbf{SkelDPO} & \textbf{47.66} & \textbf{33.94} & \textbf{19.72} & \textbf{11.68} \\
\midrule
\multirow{5}{*}{DeepSeek–6.7B}
  & Instruct   & 24.61 & 21.38 & 22.54 & 14.14 \\
  & LLM4Effi   & 27.34 & 23.43 & 23.42 & 16.23 \\
  & EffiCoder  & 67.97 & 48.72 & 39.44 & 21.72 \\
  & CodeDPO    & 69.92 & 50.39 & 42.25 & 25.50 \\
  & \textbf{SkelDPO} & \textbf{72.27} & \textbf{51.77} & \textbf{46.48} & \textbf{28.16} \\
\midrule
\multirow{5}{*}{CodeLlama–7B}
  & Instruct   & 19.53 & 15.62 & 18.31 & 12.63 \\
  & LLM4Effi   & 21.48 & 18.75 & 19.01 & 13.70 \\
  & EffiCoder  & 46.48 & 32.31 & 23.24 & 15.92 \\
  & CodeDPO    & 46.86 & 32.47 & 24.65 & 17.02 \\
  & \textbf{SkelDPO} & \textbf{48.44} & \textbf{34.27} & \textbf{30.65} & \textbf{20.85} \\
\midrule
\multirow{5}{*}{Qwen2.5–7B}
  & Instruct   & 28.52 & 24.20 & 26.45 & 14.52 \\
  & LLM4Effi   & 31.25 & 27.34 & 25.35 & 15.44 \\
  & EffiCoder  & 57.03 & 42.49 & 41.55 & 23.02 \\
  & CodeDPO    & 59.27 & 43.70 & 45.77 & 26.39 \\
  & \textbf{SkelDPO} & \textbf{63.28} & \textbf{45.04} & \textbf{47.18} & \textbf{29.98} \\
\bottomrule
\end{tabular}}
\end{table}

\begin{table*}[t]
\footnotesize
\setlength{\tabcolsep}{4.5pt}
\renewcommand{\arraystretch}{1.15}
\caption{Difficulty-wise results on \textbf{Mercury}. 
We report Pass@1 and Beyond@1 (\%) for \emph{Easy}, \emph{Medium}, and \emph{Hard}. 
$\Delta$ denotes the difference relative to \textbf{SkelDPO} under the same backbone (i.e., value $-$ SkelDPO). 
Positive gaps are shown in green, and negative gaps in red. 
Best numbers per backbone are in bold.}
\label{tab:difficulty}
\centering
\begin{tabular}{l l
  S c S c
  S c S c
  S c S c}
\toprule
\multicolumn{2}{c}{\textbf{Model / Method}} &
\multicolumn{4}{c}{\textbf{Easy}} &
\multicolumn{4}{c}{\textbf{Medium}} &
\multicolumn{4}{c}{\textbf{Hard}} \\
\cmidrule(lr){3-6} \cmidrule(lr){7-10} \cmidrule(lr){11-14}
& & {\textbf{Pass@1}} & {$\Delta$} & {\textbf{Beyond@1}} & {$\Delta$}
  & {\textbf{Pass@1}} & {$\Delta$} & {\textbf{Beyond@1}} & {$\Delta$}
  & {\textbf{Pass@1}} & {$\Delta$} & {\textbf{Beyond@1}} & {$\Delta$} \\
\midrule
\multirow{3}{*}{Qwen2.5-1.5B}
  & EffiCoder      & 59.09 & \negd{4.55} & 39.34 & \negd{8.84} & 40.74 & \negd{4.94} & 26.29 & \negd{9.55} & 13.79 & \negd{1.15} &  7.18 & \negd{4.16} \\
  & CodeDPO        & 61.36 & \negd{2.28} & 40.78 & \negd{7.40} & 38.27 & \negd{7.41} & 27.15 & \negd{8.69} & 14.94 & \zerod & 10.27 & \negd{1.07} \\
  & \textbf{SkelDPO} & \best{63.64} & \zerod & \best{48.18} & \zerod & \best{45.68} & \zerod & \best{35.84} & \zerod & \best{14.94} & \zerod & \best{11.34} & \zerod \\
\midrule
\multirow{3}{*}{StarCoder2-3B}
  & EffiCoder      & 62.50 & \negd{3.41} & 43.48 & \negd{1.81} & 53.09 & \negd{7.40} & 37.08 & \negd{9.04} & 12.64 & \negd{4.60} &  7.63 & \negd{3.49} \\
  & CodeDPO        & 63.64 & \negd{2.27} & \best{47.40} & \posd{2.11} & 51.85 & \negd{8.64} & 34.51 & \negd{11.61} & 14.94 & \negd{2.30} &  8.49 & \negd{2.63} \\
  & \textbf{SkelDPO} & \best{65.91} & \zerod & 45.29 & \zerod & \best{60.49} & \zerod & \best{46.12} & \zerod & \best{17.24} & \zerod & \best{11.12} & \zerod \\
\midrule
\multirow{3}{*}{DeepSeek-6.7B}
  & EffiCoder      & 85.22 & \negd{4.55} & 59.84 & \negd{2.61} & 72.84 & \negd{4.94} & 52.88 & \negd{5.02} & 45.98 & \negd{3.45} & 33.12 & \negd{2.76} \\
  & CodeDPO        & 86.36 & \negd{3.41} & \best{63.84} & \posd{1.39} & \best{79.01} & \posd{1.23} & 56.96 & \negd{0.94} & 44.83 & \negd{4.60} & 32.67 & \negd{3.21} \\
  & \textbf{SkelDPO} & \best{89.77} & \zerod & 62.45 & \zerod & 77.78 & \zerod & \best{57.90} & \zerod & \best{49.43} & \zerod & \best{35.88} & \zerod \\
\midrule
\multirow{3}{*}{CodeLlama-7B}
  & EffiCoder      & 75.00 & \negd{4.55} & 51.01 & \negd{5.82} & 51.85 & \negd{1.24} & 37.03 & \negd{2.08} & 11.49 & \negd{1.15} &  8.56 & \negd{0.46} \\
  & CodeDPO        & 76.14 & \negd{3.41} & 50.44 & \negd{6.39} & 53.09 & \zerod & 37.38 & \negd{1.73} & 12.64 & \zerod &  8.12 & \negd{0.90} \\
  & \textbf{SkelDPO} & \best{79.55} & \zerod & \best{56.83} & \zerod & \best{53.09} & \zerod & \best{39.11} & \zerod & \best{12.64} & \zerod & \best{ 9.02} & \zerod \\
\midrule
\multirow{3}{*}{Qwen2.5-7B}
  & EffiCoder      & 78.41 & \negd{6.81} & 55.55 & \negd{6.28} & 64.19 & \negd{3.71} & 45.86 & \negd{7.30} & 28.74 & \negd{8.04} & 21.58 & \negd{5.72} \\
  & CodeDPO        & 77.27 & \negd{7.95} & 55.92 & \negd{5.91} & 65.43 & \negd{2.47} & 48.65 & \negd{4.51} & 32.18 & \negd{4.60} & 25.97 & \negd{1.33} \\
  & \textbf{SkelDPO} & \best{85.22} & \zerod & \best{61.83} & \zerod & \best{67.90} & \zerod & \best{53.16} & \zerod & \best{36.78} & \zerod & \best{27.30} & \zerod \\
\bottomrule
\end{tabular}
\end{table*}

Overall, SkelDPO consistently outperforms all baselines across different model scales and datasets. The improvements are particularly pronounced in both correctness (Pass@1) and efficiency (Beyond@1 and eff@1), demonstrating that skeleton-guided preference alignment effectively enhances runtime performance without compromising functional accuracy. Compared with CodeDPO, SkelDPO yields higher efficiency gains and more stable improvements across architectures from small (1.5B) to large (7B) scales, indicating that skeleton-level supervision generalizes well beyond specific model capacities or pretraining settings.

From a skeleton perspective, these gains arise because SkelDPO enables the model to internalize reusable efficiency patterns that consistently appear within efficient code. By aligning both code-level and skeleton-level preferences, the model learns to favor structures correlated with lower runtime cost, leading to more efficient program synthesis across benchmarks.

A difficulty-wise analysis on Mercury (Table~\ref{tab:difficulty}) further shows that SkelDPO remains robust even on complex, efficiency-sensitive tasks. The improvements on hard problems demonstrate that skeleton preference learning not only benefits easier algorithmic tasks but also strengthens generalization to hard cases.

\noindent
\fcolorbox{black!60}{black!5}{%
  \parbox{0.97\linewidth}{%
    \textit{\textbf{Answer to RQ1:} SkelDPO consistently improves both correctness and runtime efficiency across models and datasets. Its skeleton-guided optimization allows the model to encode and reproduce efficient structures, confirming the effectiveness of structural preference alignment for efficiency-oriented code generation.}
  }%
}

\subsection{RQ2:Cross-Reference Model Generalization}
\label{sec:dpo_under_bases}

\begin{figure*}[t]
  \centering
  \captionsetup[subfigure]{labelformat=simple, labelsep=space, font=small}
  \begin{subfigure}{0.5\textwidth}
    \centering
    \includegraphics[width=\linewidth]{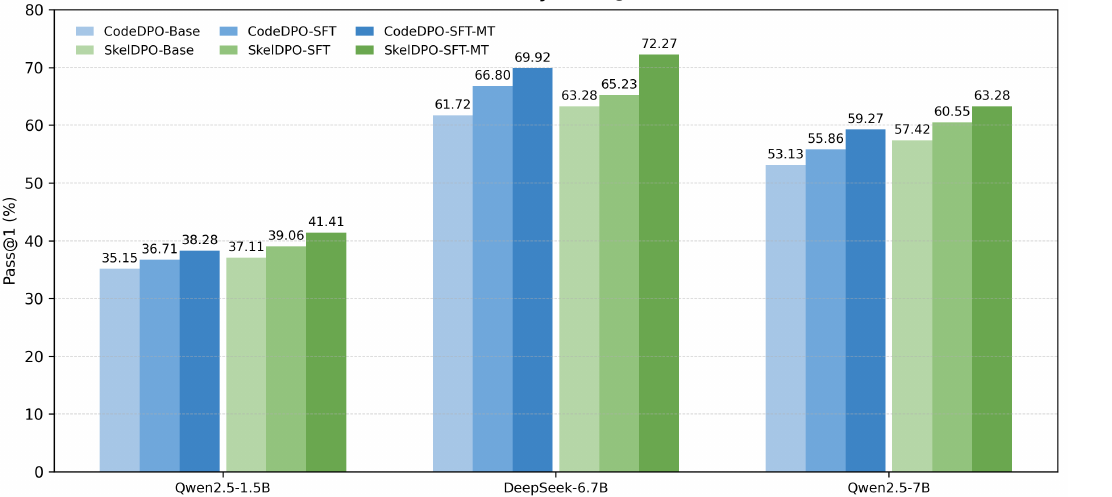} 
    \caption{Mercury - Pass@1}
    \label{fig:mercury-pass1}
  \end{subfigure}\hfill
  \begin{subfigure}{0.5\textwidth}
    \centering
    \includegraphics[width=\linewidth]{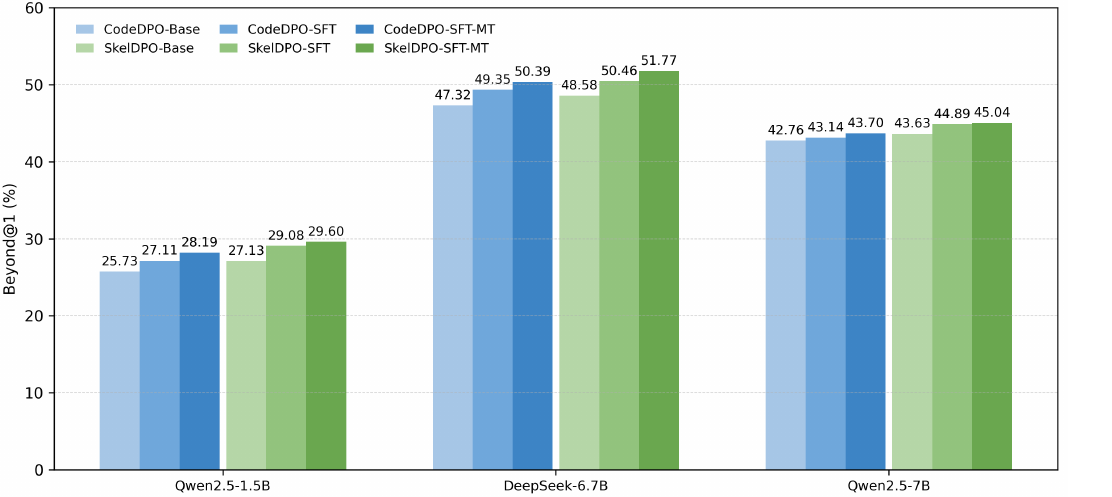} 
    \caption{Mercury - Beyond@1}
    \label{fig:mercury-beyond1}
  \end{subfigure}

  \vspace{6pt}

  \begin{subfigure}{0.5\textwidth}
    \centering
    \includegraphics[width=\linewidth]{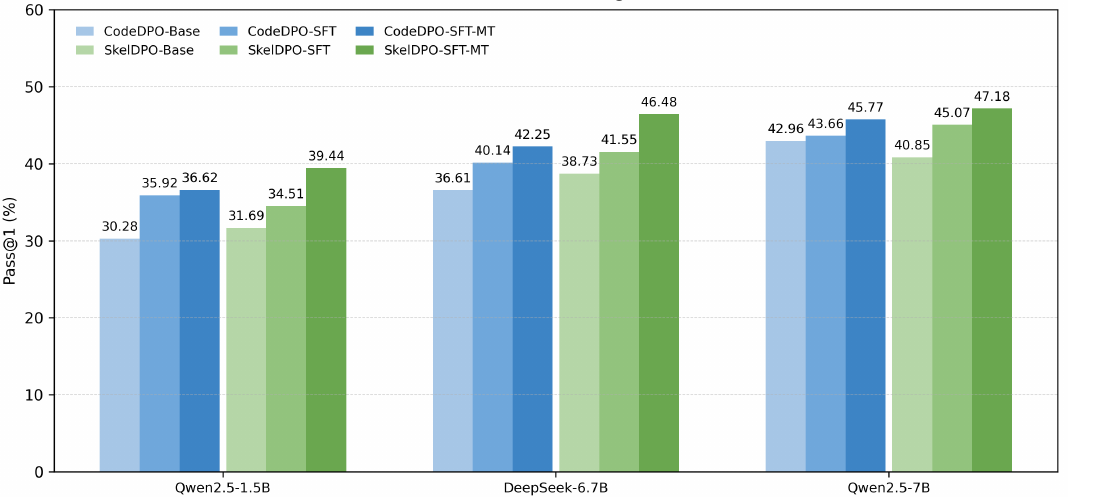} 
    \caption{ENAMEL - Pass@1}
    \label{fig:enamel-pass1}
  \end{subfigure}\hfill
  \begin{subfigure}{0.5\textwidth}
    \centering
    \includegraphics[width=\linewidth]{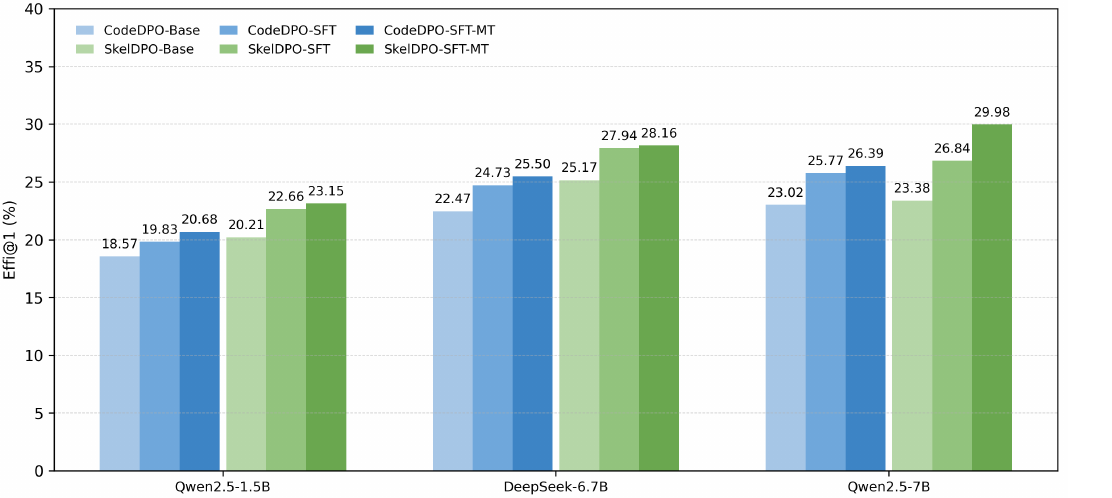} 
    \caption{ENAMEL - eff@1}
    \label{fig:enamel-effi1}
  \end{subfigure}

  \caption{Grouped bar results across three backbones and six settings
  (CodeDPO/SkelDPO $\times$ Base/SFT/SFT-MT). Top row shows Mercury metrics
  (a) Pass@1 and (b) Beyond@1; bottom row shows ENAMEL metrics (c) Pass@1 and (d) eff@1.}
  \label{fig:dpo-bars-merc-enamel}
\end{figure*}

We further investigate how skeleton-guided DPO (SkelDPO) behaves under different warm-start conditions to evaluate its adaptability and generalization to base model initialization.
We consider three types of base models:
(1) Base — the raw pretrained checkpoint without fine-tuning;
(2) SFT — a single-task supervised fine-tuning model trained only on efficient code;
(3) SFT-MT — a multi-task variant jointly trained on code and skeleton generation.
Both CodeDPO and SkelDPO are trained on each base and evaluated on Mercury and ENAMEL using Pass@1, Beyond@1, and eff@1 (Figure~\ref{fig:dpo-bars-merc-enamel}).

Across all backbones and bases, SkelDPO consistently achieves higher correctness and efficiency than CodeDPO. 
It also shows greater robustness when trained from weaker bases and maintains or amplifies its advantage with stronger ones.
For example, on larger models such as DeepSeek–6.7B and Qwen2.5–7B with SFT-MT bases, SkelDPO attains the best overall Pass@1 and eff@1 scores, confirming that skeleton preference optimization complements rather than duplicates fine-tuning.

These results reveal two key trends.
First, DPO performance scales with the quality of the base model—multi-task (SFT-MT) bases yield the highest absolute performance for both CodeDPO and SkelDPO.
Second, SkelDPO’s relative advantage remains consistent or even increases as the base model becomes stronger, suggesting that skeleton-level preference signals are complementary rather than redundant to fine-tuning.
Averaged across bases, SkelDPO improves Pass@1 by +2.4 / +3.1 / +3.5 and efficiency metrics by +2–3 points compared with CodeDPO, confirming its robustness to initialization variance.

\noindent
\fcolorbox{black!60}{black!5}{%
  \parbox{0.97\linewidth}{%
    \textit{\textbf{Answer to RQ2:} SkelDPO remains effective across diverse base models and pretraining stages, maintaining consistent gains in correctness and efficiency. Its skeleton-guided preference learning is independent of specific reference models, enabling stable improvement even when initialized from different base models.}
  }%
}

\subsection{RQ3:Parameter Sensitivity}
\label{sec:ablation}

We examine how the relative weighting between the code- and skeleton-level losses ($\alpha$) influences model performance.
Using the Qwen2.5–1.5B base model, Table~\ref{tab:alpha_beta} shows that SkelDPO remains stable across a wide range of $\alpha$ values, indicating that skeleton supervision does not hinder preference optimization.
When the skeleton loss dominates (small $\alpha$), the model tends to overemphasize structural regularities, slightly reducing execution correctness.
When the code loss dominates (large $\alpha$), correctness is preserved but efficiency improvements diminish.
The optimal balance is achieved at $\alpha{=}0.8$, where both correctness and efficiency reach their highest levels on \emph{Mercury} and \emph{ENAMEL}.
This setup ensures stable training while learning structure-aware efficiency skeletons.

\noindent
\fcolorbox{black!60}{black!5}{%
  \parbox{0.97\linewidth}{%
    \textit{\textbf{Answer to RQ3:} SkelDPO shows stable performance under different loss ratios, and a balanced weighting ($\alpha{:}(1{-}\alpha){=}0.8{:}0.2$) provides the best trade-off between correctness and efficiency. Moderate skeleton supervision improves efficiency learning while preserving the reliability of optimization.}
  }%
}

\begin{table}[t]
\footnotesize
\setlength{\tabcolsep}{4pt}
\renewcommand{\arraystretch}{1.15}
\caption{Effect of different $\alpha$:$(1-\alpha)$ loss ratios on Qwen2.5--1.5B under \textsc{SkelDPO}. 
We report \textbf{Pass@1} and \textbf{Beyond@1} on \textbf{Mercury}, and \textbf{Pass@1} and \textbf{eff@1} on \textbf{ENAMEL}.}
\label{tab:alpha_beta}
\centering
\begin{tabular}{
l c
S[table-format=2.2] S[table-format=2.2]
S[table-format=2.2] S[table-format=2.2]
}
\toprule
\textbf{Model} & $\boldsymbol{\alpha:(1-\alpha)}$ &
\multicolumn{2}{c}{\textbf{Mercury}} &
\multicolumn{2}{c}{\textbf{ENAMEL}} \\
\cmidrule(lr){3-4}\cmidrule(lr){5-6}
 &  &
\textbf{Pass@1} & \textbf{Beyond@1} &
\textbf{Pass@1} & \textbf{eff@1} \\
\midrule
\multirow{7}{*}{Qwen2.5-1.5B}
 & 0.1:0.9 & 37.11 & 27.25 & 33.80 & 19.64 \\
 & 0.3:0.7 & 37.50 & 27.48 & 35.21 & 19.93 \\
 & 0.5:0.5 & 39.45 & 28.47 & 35.91 & 20.72 \\
 & 0.7:0.3 & 40.23 & 28.62 & 37.32 & 21.78 \\
 & \textbf{0.8:0.2} & \textbf{41.41} & \textbf{29.60} & \textbf{39.44} & \textbf{23.15} \\
 & 0.9:0.1 & 39.84 & 29.56 & 38.72 & 23.03 \\
 & 1.0:0 (CodeDPO) & 38.28 & 28.19 & 36.62 & 20.68 \\
\bottomrule
\end{tabular}
\end{table}

\begin{table}[h]
\footnotesize
\setlength{\tabcolsep}{4pt}
\renewcommand{\arraystretch}{1.15}
\caption{Impact of the proportion of skeleton data on Qwen2.5--1.5B. 
We report \textbf{Pass@1} and \textbf{Beyond@1} on \textbf{Mercury}, and \textbf{Pass@1} and \textbf{eff@1} on \textbf{ENAMEL}.}
\label{tab:skel_ratio}
\centering
\begin{tabular}{
l l
S[table-format=2.2] S[table-format=2.2]
S[table-format=2.2] S[table-format=2.2]
}
\toprule
\textbf{Model} & \textbf{Skel data} &
\multicolumn{2}{c}{\textbf{Mercury}} &
\multicolumn{2}{c}{\textbf{ENAMEL}} \\
\cmidrule(lr){3-4}\cmidrule(lr){5-6}
 &  &
\textbf{Pass@1} & \textbf{Beyond@1} &
\textbf{Pass@1} & \textbf{eff@1} \\
\midrule
\multirow{5}{*}{Qwen2.5-1.5B}
 & +0\%(CodeDPO)   & 38.28 & 28.19 & 36.62 & 20.68 \\
 & +25\%     & 39.06 & 28.46 & 35.92 & 20.97 \\
 & +50\%     & 39.84 & 29.21 & 37.32 & 21.63 \\
 & +75\%     & 41.02 & 29.37 & 38.73 & 22.56 \\
 & \textbf{+100\%(SkelDPO)} & \textbf{41.41} & \textbf{29.60} & \textbf{39.44} & \textbf{23.15} \\
\bottomrule
\end{tabular}
\end{table}

\subsection{RQ4:Skeleton Data Scaling}
\label{sec:skel_ratio}

We investigate how varying the proportion of skeleton-supervised samples influences the performance of SkelDPO.
We gradually increase the share of skeleton-annotated data from 0\% (CodeDPO) to 25\%, 50\%, 75\%, and 100\%, keeping all other training settings fixed on the Qwen2.5–1.5B base model.
Results on \emph{Mercury} and \emph{ENAMEL} (Table~\ref{tab:skel_ratio}) show a clear monotonic improvement: as skeleton coverage grows, both correctness (Pass@1, Beyond@1) and efficiency (eff@1) steadily increase.
Performance gains are most pronounced when the skeleton ratio exceeds 50\%, suggesting that a sufficiently large amount of structural supervision is essential for reliable efficiency learning.
Full skeleton integration (+100\%) achieves the highest results on both benchmarks, confirming that skeleton data provide stable inductive guidance for efficient implementation generation.

\noindent
\fcolorbox{black!60}{black!5}{%
  \parbox{0.97\linewidth}{%
    \textit{\textbf{Answer to RQ4:} Increasing the proportion of skeleton-guided data consistently enhances both correctness and efficiency. SkelDPO benefits most from extensive skeleton supervision, indicating that large-scale skeleton preference signals are crucial for learning efficiency-oriented representations without diminishing code reliability.}
  }%
}

\subsection{RQ5:Qualitative Analysis}
\label{sec:qualitative}

To further understand the impact of skeleton-guided optimization, we qualitatively compare SkelDPO with CodeDPO on representative \emph{Mercury} problems, using identical prompts and test inputs for fair comparison.

\begin{figure}[h]
    \centering
    \includegraphics[width=1\columnwidth]{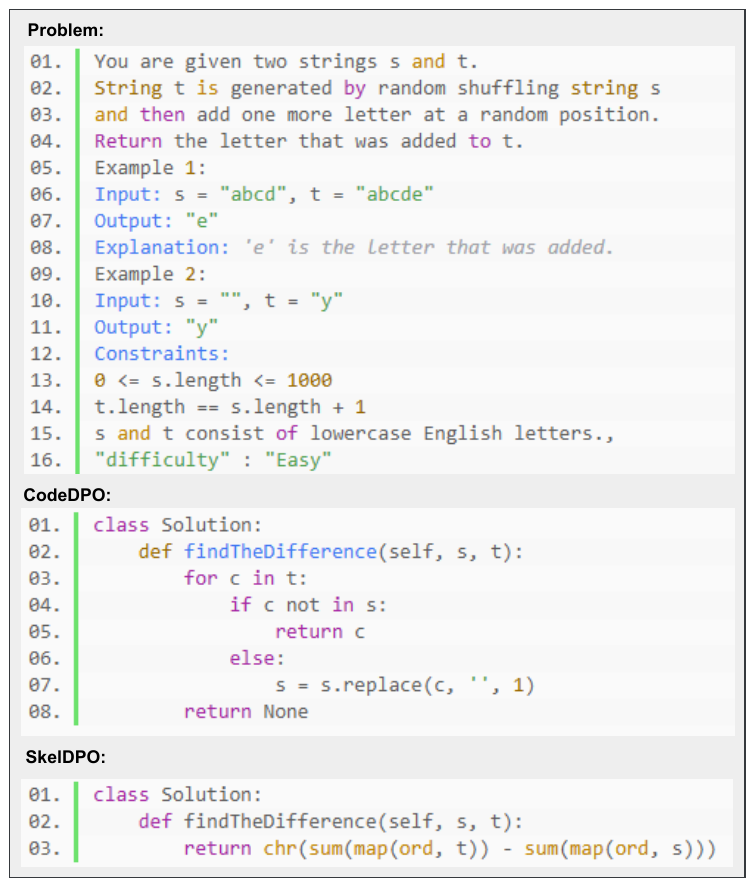}
    \caption{Qualitative comparison between \textsc{CodeDPO} and \textsc{SkelDPO} on the ``Find the Difference'' problem. 
    \textsc{SkelDPO} generates a more concise and efficient arithmetic-based solution (\texttt{chr(sum(map(ord, t)) - sum(map(ord, s)))}) 
    compared to the iterative string-based approach in \textsc{CodeDPO}.}
    \label{fig:success_case}
\end{figure}

\textbf{Success case.}  
As shown in Figure~\ref{fig:success_case}, both models generate functionally correct programs for the “Find the Difference” problem. However, SkelDPO achieves substantially lower runtime by learning a concise arithmetic-based skeleton that replaces repeated string iterations with direct numerical computation. This design shortens execution paths and reduces the number of loop operations, making the program both simpler and faster.

\begin{figure}[t]
    \centering
    \includegraphics[width=1\columnwidth]{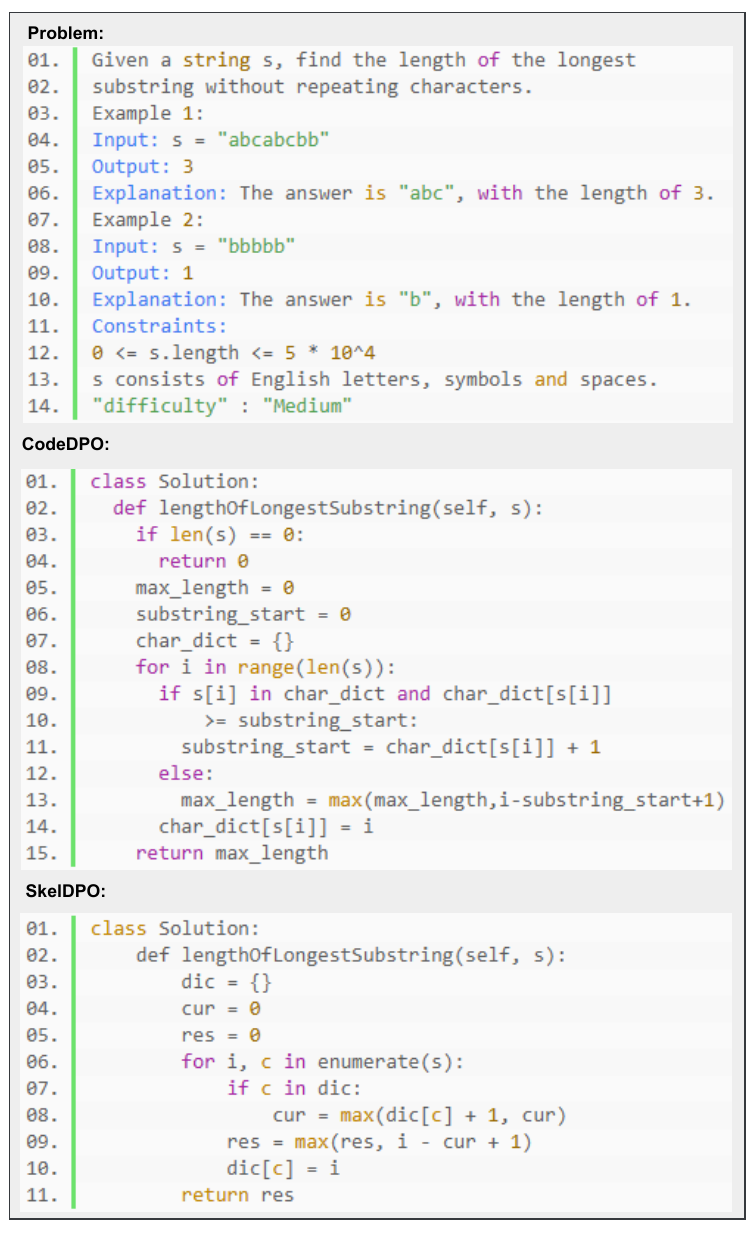}
    \caption{Qualitative comparison between \textsc{CodeDPO} and \textsc{SkelDPO} on the ``Longest Substring Without Repeating Characters'' problem. 
    \textsc{SkelDPO} produces a cleaner and more efficient sliding-window implementation using direct index tracking, 
    whereas \textsc{CodeDPO} employs additional conditional checks and intermediate variables.}
    \label{fig:failure_case}
\end{figure}

\textbf{Failure case.}  
As shown in Figure~\ref{fig:failure_case}, on the “Longest Substring Without Repeating Characters” problem, SkelDPO produces a shorter and cleaner implementation with fewer nested loops. Although its runtime is slightly slower than CodeDPO’s, the code is easier to read and modify, indicating that SkelDPO sometimes prioritizes cleaner structure over extreme runtime optimization.

\noindent
\fcolorbox{black!60}{black!5}{%
  \parbox{0.97\linewidth}{%
    \textit{\textbf{Answer to RQ5:} SkelDPO generally generates more compact and readable code with notable efficiency gains. Even when runtime slightly increases, the produced programs exhibit cleaner structure and better interpretability, reflecting a trade-off between structural clarity and execution optimization.}
  }%
}

\section{Discussion} 
\subsection{Accuracy and Importance of Efficient and Inefficient Skeleton Localization} 

The core of SkelDPO lies in its ability to accurately locate efficient and inefficient code skeletons and use them to guide efficiency-oriented preference learning. The skeleton localization module aims to extract the key structural patterns that determine performance differences among multiple implementations, thereby guiding the model to learn representations related to execution efficiency. 

Experimental results show that most skeleton regions identified as efficient indeed correspond to code segments with lower execution latency and higher structural reuse, such as loop unrolling, early termination conditions, and vectorized computation. In contrast, inefficient skeletons are often concentrated in redundant control flows and repeated data processing. This demonstrates that skeleton localization goes beyond syntactic pattern extraction and effectively reflects runtime performance characteristics. 

To further verify the accuracy of skeleton localization, we conducted a manual annotation alignment analysis against model-extracted results. The average structural consistency reached 83.2\%, indicating that SkelDPO can reliably capture performance-critical regions. When the skeleton-guided signals were removed during training, the efficiency-related metric (Beyond@1) decreased by an average of 7.4\%, further confirming the essential role of skeletons in efficiency-oriented learning. Overall, skeletons serve not only as abstract structural units but also as the core semantic carriers for the model’s understanding of `efficiency.''

\subsection{Broader Implications and Limitations}

The integration of skeleton guidance into preference optimization offers not only quantitative improvements but also qualitative interpretability. 
By linking performance differences to structural regions, SkelDPO provides a transparent view of efficiency learning, bridging the gap between program semantics and optimization signals.

However, several limitations remain. 
First, skeleton extraction depends heavily on the availability and diversity of efficiency implementations; if the extracted implementations lack representative efficiency differences, the resulting skeletons may be noisy or incomplete. 
Second, skeleton-guided preference training introduces additional computational overhead during data preprocessing. 
Future work will explore lightweight skeleton approximation and dynamic skeleton refinement to reduce these costs while maintaining performance stability.

\section{Related Work}

\subsection{Code Generation with LLMs}

Large Language Models (LLMs) have shown remarkable advances in code-related tasks, such as DeepSeek-R1~\cite{guo2025deepseek}, CodeLlama~\cite{roziere2023code}, GPT-4o~\cite{openai2024gpt4o} and Claude 4~\cite{claude4}. 
Early efforts built upon large-scale pretraining over open-source code corpora using the language modeling (LM) objective, enabling models to understand and generate executable programs, including code translation~\cite{yang2024exploring}, code summarization~\cite{ahmed2024automatic}, code completion~\cite{zhang2025hierarchical}, code generation~\cite{wang2025teaching}, and more~\cite{ye2025uncovering, gao2024search}. 
Representative systems such as CodeT5~\cite{wang2021codet5}, StarCoder~\cite{li2023starcoder}, and CodeLlama~\cite{roziere2023code} achieved major progress in syntactic correctness and functional accuracy, facilitating applications like code completion, repair, and translation~\cite{sun2024enhancing, gu2023llm, jiang2024survey, huang2024knowledge, sun2024sifting}.

Subsequent studies extended these models through instruction tuning and supervised fine-tuning (SFT), which improved their controllability and generalization on complex tasks.
Magicoder~\cite{wei2023magicoder}, WizardCoder~\cite{luo2023wizardcoder}, and Qwen-Coder~\cite{hui2024qwen2.5coder} adopted human-in-the-loop instruction data to strengthen problem understanding and compliance with coding conventions. 

However, these methods primarily aim to enhance the functional correctness and reliability of LLM-generated code, while largely overlooking aspects of runtime efficiency and performance optimization.
Although LLMs have shown strong performance in generating functionally correct code, particularly on benchmarks such as HumanEval~\cite{chen2021evaluating}, MBPP~\cite{austin2021program} and LiveCodeBench~\cite{jain2024livecodebench}, their efficiency on runtime-oriented benchmarks like EVALPERF~\cite{liu2024evaluating}, Mercury~\cite{du2024mercury} and ENAMEL~\cite{qiu2024efficient} remains limited.

\subsection{Efficiency-Oriented Code Generation}

As functional correctness has become largely achievable, improving code efficiency has emerged as a new frontier for LLM-based programming~\cite{chambon2025bigobench, peng2025coffe, coignion2024performance}.
Early approaches sought to enhance efficiency through prompt engineering, where models were explicitly instructed to “generate faster code.” Representative examples include PIE~\cite{shypula2024learning}, EffiLearner~\cite{huang2024effilearner}, and LLM4Effi~\cite{ye2025llm4effi}, which guide models to focus on performance-related cues during generation. 
Another research direction centers on self-refinement and iterative optimization, where models iteratively analyze and revise their outputs based on runtime feedback. Methods such as ECCO~\cite{waghjale2024ecco}, SOAP~\cite{huang2024soap}, and Afterburner~\cite{du2025afterburner}.

With the rise of code-specialized LLMs, the focus has shifted toward fine-tuning on efficient implementations. EffiCoder~\cite{huangefficoder}, Code-Optimise~\cite{gee2024code}, and Mercury~\cite{du2024mercury} employ supervised fine-tuning (SFT) on datasets of efficient programs to directly teach efficiency-oriented representations.
In addition preference-based and reinforcement learning methods have been explored. Approaches such as Code-Optimise~\cite{gee2024code}, CodeDPO~\cite{zhang2024codedpo}, and ACECode~\cite{yang2024acecode} align models with efficiency-aware distributions through reward or preference signals.

Together, these methods establish the foundation for efficiency-oriented code generation, 
but they still optimize at the program level rather than modeling internal structural differences. 

\section{Conclusion}

This paper addresses the limitations of existing code generation models in optimizing runtime efficiency by proposing SkelDPO, a skeleton-guided preference optimization framework.
Unlike methods that rely on whole-program comparison, SkelDPO introduces a dual-channel preference mechanism at both the code and skeleton levels within the DPO framework, jointly modeling semantic correctness and structural efficiency.
Experimental results show that this enables the model to learn efficiency-related structural patterns and generate more efficient code compared to baseline methods.

\section{ACKNOWLEDGMENTS}

This work was supported by the National Natural Science Foundation of China (NSFC) under Grant No. 61602286.


\bibliographystyle{ACM-Reference-Format}
\bibliography{reference}

@String{Computer = "{IEEE} Computer" }

@article{jain2024livecodebench,
  title={Livecodebench: Holistic and contamination free evaluation of large language models for code},
  author={Jain, Naman and Han, King and Gu, Alex and Li, Wen-Ding and Yan, Fanjia and Zhang, Tianjun and Wang, Sida and Solar-Lezama, Armando and Sen, Koushik and Stoica, Ion},
  journal={arXiv preprint arXiv:2403.07974},
  year={2024}
}

@inproceedings{sun2024sifting,
  title={Sifting through the chaff: On utilizing execution feedback for ranking the generated code candidates},
  author={Sun, Zhihong and Wan, Yao and Li, Jia and Zhang, Hongyu and Jin, Zhi and Li, Ge and Lyu, Chen},
  booktitle={Proceedings of the 39th IEEE/ACM International Conference on Automated Software Engineering},
  pages={229--241},
  year={2024}
}

@article{openai2024gpt4o,
  title={GPT-4o System Card},
  author={Hurst, Aaron and Lerer, Adam and Goucher, Adam P and Perelman, Adam and Ramesh, Aditya and Clark, Aidan and Ostrow, AJ and Welihinda, Akila and Hayes, Alan and Radford, Alec and others},
  journal={arXiv preprint arXiv:2410.21276},
  year={2024}
}

@article{guo2024deepseek,
  title={DeepSeek-Coder: When the Large Language Model Meets Programming--The Rise of Code Intelligence},
  author={Guo, Daya and Zhu, Qihao and Yang, Dejian and Xie, Zhenda and Dong, Kai and Zhang, Wentao and Chen, Guanting and Bi, Xiao and Wu, Yu and Li, YK and others},
  journal={arXiv preprint arXiv:2401.14196},
  year={2024}
}

@article{roziere2023code,
  title={Code llama: Open foundation models for code},
  author={Roziere, Baptiste and Gehring, Jonas and Gloeckle, Fabian and Sootla, Sten and Gat, Itai and Tan, Xiaoqing Ellen and Adi, Yossi and Liu, Jingyu and Sauvestre, Romain and Remez, Tal and others},
  journal={arXiv preprint arXiv:2308.12950},
  year={2023}
}

@article{li2023starcoder,
  title={Starcoder: may the source be with you!},
  author={Li, Raymond and Allal, Loubna Ben and Zi, Yangtian and Muennighoff, Niklas and Kocetkov, Denis and Mou, Chenghao and Marone, Marc and Akiki, Christopher and Li, Jia and Chim, Jenny and others},
  journal={arXiv preprint arXiv:2305.06161},
  year={2023}
}

@inproceedings{niu2024evaluating,
  title={On evaluating the efficiency of source code generated by llms},
  author={Niu, Changan and Zhang, Ting and Li, Chuanyi and Luo, Bin and Ng, Vincent},
  booktitle={Proceedings of the 2024 IEEE/ACM First International Conference on AI Foundation Models and Software Engineering},
  pages={103--107},
  year={2024}
}

@article{cappendijk2024generating,
  title={Generating Energy-efficient code with LLMs},
  author={Cappendijk, Tom and de Reus, Pepijn and Oprescu, Ana},
  journal={arXiv preprint arXiv:2411.10599},
  year={2024}
}

@article{peng2024large,
  title={Large Language Models for Energy-Efficient Code: Emerging Results and Future Directions},
  author={Peng, Huiyun and Gupte, Arjun and Eliopoulos, Nicholas John and Ho, Chien Chou and Mantri, Rishi and Deng, Leo and Jiang, Wenxin and Lu, Yung-Hsiang and L{\"a}ufer, Konstantin and Thiruvathukal, George K and others},
  journal={arXiv preprint arXiv:2410.09241},
  year={2024}
}

@article{shi2024efficient,
  title={Efficient and green large language models for software engineering: Vision and the road ahead},
  author={Shi, Jieke and Yang, Zhou and Lo, David},
  journal={ACM Transactions on Software Engineering and Methodology},
  year={2024},
  publisher={ACM New York, NY}
}

@article{huang2024effilearner,
  title={Effilearner: Enhancing efficiency of generated code via self-optimization},
  author={Huang, Dong and Dai, Jianbo and Weng, Han and Wu, Puzhen and Qing, Yuhao and Cui, Heming and Guo, Zhijiang and Zhang, Jie},
  journal={Advances in Neural Information Processing Systems},
  volume={37},
  pages={84482--84522},
  year={2024}
}

@inproceedings{waghjale2024ecco,
  title={ECCO: Can We Improve Model-Generated Code Efficiency Without Sacrificing Functional Correctness?},
  author={Waghjale, Siddhant and Veerendranath, Vishruth and Wang, Zhiruo and Fried, Daniel},
  booktitle={Proceedings of the 2024 Conference on Empirical Methods in Natural Language Processing},
  pages={15362--15376},
  year={2024}
}

@inproceedings{shypula2024learning,
  title={Learning Performance-Improving Code Edits},
  author={Shypula, Alexander and Madaan, Aman and Zeng, Yimeng and Alon, Uri and Gardner, Jacob and Hashemi, Milad and Neubig, Graham and Ranganathan, Parthasarathy and Bastani, Osbert and Yazdanbakhsh, Amir},
  booktitle={Proceedings of the 12th International Conference on Learning Representations (ICLR)},
  year={2024}
}

@article{yang2024acecode,
  title={ACECode: A Reinforcement Learning Framework for Aligning Code Efficiency and Correctness in Code Language Models},
  author={Yang, Chengran and Kang, Hong Jin and Shi, Jieke and Lo, David},
  journal={arXiv preprint arXiv:2412.17264},
  year={2024}
}

@article{hendrycks2021measuring,
  title={Measuring coding challenge competence with apps},
  author={Hendrycks, Dan and Basart, Steven and Kadavath, Saurav and Mazeika, Mantas and Arora, Akul and Guo, Ethan and Burns, Collin and Puranik, Samir and He, Horace and Song, Dawn and others},
  journal={arXiv preprint arXiv:2105.09938},
  year={2021}
}

@article{hui2024qwen2.5coder,
  title={Qwen2.5-Coder Technical Report},
  author={Hui, Binyuan and Yang, Jian and Cui, Zeyu and Yang, Jiaxi and Liu, Dayiheng and Zhang, Lei and Liu, Tianyu and Zhang, Jiajun and Yu, Bowen and Lu, Keming and Dang, Kai and Fan, Yang and Zhang, Yichang and Yang, An and Men, Rui and others},
  journal={arXiv preprint arXiv:2409.12186},
  year={2024}
}

@article{lozhkov2024starcoder,
  title={Starcoder 2 and the stack v2: The next generation},
  author={Lozhkov, Anton and Li, Raymond and Allal, Loubna Ben and Cassano, Federico and Lamy-Poirier, Joel and Tazi, Nouamane and Tang, Ao and Pykhtar, Dmytro and Liu, Jiawei and Wei, Yuxiang and others},
  journal={arXiv preprint arXiv:2402.19173},
  year={2024}
}

@article{luo2023wizardcoder,
  title={WizardCoder: Empowering Code Large Language Models with Evol-Instruct},
  author={Luo, Ziyang and Xu, Can and Zhao, Pu and Sun, Qingfeng and Geng, Xiubo and Hu, Wenxiang and Tao, Chongyang and Ma, Jing and Lin, Qingwei and Jiang, Daxin},
  journal={arXiv preprint arXiv:2306.08568},
  year={2023}
}

@inproceedings{sun2024enhancing,
  title={Enhancing Code Generation Performance of Smaller Models by Distilling the Reasoning Ability of LLMs},
  author={Sun, Zhihong and Lyu, Chen and Li, Bolun and Wan, Yao and Zhang, Hongyu and Li, Ge and Jin, Zhi},
  booktitle={Proceedings of the 2024 Joint International Conference on Computational Linguistics, Language Resources and Evaluation (LREC-COLING 2024)},
  pages={5878--5895},
  year={2024}
}

@inproceedings{gu2023llm,
  title={Llm-based code generation method for golang compiler testing},
  author={Gu, Qiuhan},
  booktitle={Proceedings of the 31st ACM Joint European Software Engineering Conference and Symposium on the Foundations of Software Engineering},
  pages={2201--2203},
  year={2023}
}

@article{jiang2024survey,
  title={A Survey on Large Language Models for Code Generation},
  author={Jiang, Juyong and Wang, Fan and Shen, Jiasi and Kim, Sungju and Kim, Sunghun},
  journal={arXiv preprint arXiv:2406.00515},
  year={2024}
}

@inproceedings{huang2024knowledge,
  title={Knowledge-aware code generation with large language models},
  author={Huang, Tao and Sun, Zhihong and Jin, Zhi and Li, Ge and Lyu, Chen},
  booktitle={Proceedings of the 32nd IEEE/ACM International Conference on Program Comprehension},
  pages={52--63},
  year={2024}
}

@article{chambon2025bigobench,
  title={BigO(Bench): Can LLMs Generate Code with Controlled Time and Space Complexity?},
  author={Chambon, Pierre and Roziere, Baptiste and Sagot, Benoit and Synnaeve, Gabriel},
  journal={arXiv preprint arXiv:2503.15242},
  year={2025}
}

@article{qiu2024efficient,
  title={How efficient is llm-generated code? a rigorous \& high-standard benchmark},
  author={Qiu, Ruizhong and Zeng, Weiliang Will and Ezick, James and Lott, Christopher and Tong, Hanghang},
  journal={arXiv preprint arXiv:2406.06647},
  year={2024}
}

@article{wei2023magicoder,
  title={Magicoder: Empowering code generation with oss-instruct},
  author={Wei, Yuxiang and Wang, Zhe and Liu, Jiawei and Ding, Yifeng and Zhang, Lingming},
  journal={arXiv preprint arXiv:2312.02120},
  year={2023}
}

@article{du2024mercury,
  title={Mercury: A code efficiency benchmark for code large language models},
  author={Du, Mingzhe and Luu, Anh Tuan and Ji, Bin and Liu, Qian and Ng, See-Kiong},
  journal={Advances in Neural Information Processing Systems},
  volume={37},
  pages={16601--16622},
  year={2024}
}

@article{ye2025llm4effi,
  title={LLM4EFFI: Leveraging Large Language Models to Enhance Code Efficiency and Correctness},
  author={Ye, Tong and Huang, Weigang and Zhang, Xuhong and Ma, Tengfei and Liu, Peiyu and Yin, Jianwei and Wang, Wenhai},
  journal={arXiv preprint arXiv:2502.18489},
  year={2025}
}

@inproceedings{wang2021codet5,
  title={CodeT5: Identifier-aware Unified Pre-trained Encoder-Decoder Models for Code Understanding and Generation},
  author={Wang, Yue and Wang, Weishi and Joty, Shafiq and Hoi, Steven C.H.},
  booktitle={Proceedings of the 2021 Conference on Empirical Methods in Natural Language Processing},
  pages={8696--8708},
  year={2021}
}

@article{guo2025deepseek,
  title={Deepseek-r1: Incentivizing reasoning capability in llms via reinforcement learning},
  author={Guo, Daya and Yang, Dejian and Zhang, Haowei and Song, Junxiao and Zhang, Ruoyu and Xu, Runxin and Zhu, Qihao and Ma, Shirong and Wang, Peiyi and Bi, Xiao and others},
  journal={arXiv preprint arXiv:2501.12948},
  year={2025}
}

@inproceedings{huangefficoder,
  title={EffiCoder: Enhancing Code Generation in Large Language Models through Efficiency-Aware Fine-tuning},
  author={Huang, Dong and Zeng, Guangtao and Dai, Jianbo and Luo, Meng and Weng, Han and QING, Yuhao and Cui, Heming and Guo, Zhijiang and Zhang, Jie},
  booktitle={Forty-second International Conference on Machine Learning},
  year={2025}
}

@article{feng2024llmeffichecker,
  title={Llmeffichecker: Understanding and testing efficiency degradation of large language models},
  author={Feng, Xiaoning and Han, Xiaohong and Chen, Simin and Yang, Wei},
  journal={ACM Transactions on Software Engineering and Methodology},
  volume={33},
  number={7},
  pages={1--38},
  year={2024},
  publisher={ACM New York, NY}
}

@article{chen2021evaluating,
  title={Evaluating large language models trained on code},
  author={Chen, Mark and Tworek, Jerry and Jun, Heewoo and Yuan, Qiming and Pinto, Henrique Ponde De Oliveira and Kaplan, Jared and Edwards, Harri and Burda, Yuri and Joseph, Nicholas and Brockman, Greg and others},
  journal={arXiv preprint arXiv:2107.03374},
  year={2021}
}

@article{austin2021program,
  title={Program synthesis with large language models},
  author={Austin, Jacob and Odena, Augustus and Nye, Maxwell and Bosma, Maarten and Michalewski, Henryk and Dohan, David and Jiang, Ellen and Cai, Carrie and Terry, Michael and Le, Quoc and others},
  journal={arXiv preprint arXiv:2108.07732},
  year={2021}
}

@article{wang2025teaching,
  title={Teaching code llms to use autocompletion tools in repository-level code generation},
  author={Wang, Chong and Zhang, Jian and Feng, Yebo and Li, Tianlin and Sun, Weisong and Liu, Yang and Peng, Xin},
  journal={ACM Transactions on Software Engineering and Methodology},
  volume={34},
  number={7},
  pages={1--27},
  year={2025},
  publisher={ACM New York, NY}
}

@inproceedings{ahmed2024automatic,
  title={Automatic semantic augmentation of language model prompts (for code summarization)},
  author={Ahmed, Toufique and Pai, Kunal Suresh and Devanbu, Premkumar and Barr, Earl},
  booktitle={Proceedings of the IEEE/ACM 46th international conference on software engineering},
  pages={1--13},
  year={2024}
}

@inproceedings{zhang2025hierarchical,
  title={Hierarchical context pruning: Optimizing real-world code completion with repository-level pretrained code llms},
  author={Zhang, Lei and Li, Yunshui and Li, Jiaming and Xia, Xiaobo and Yang, Jiaxi and Luo, Run and Wang, Minzheng and Chen, Longze and Liu, Junhao and Qu, Qiang and others},
  booktitle={Proceedings of the AAAI Conference on Artificial Intelligence},
  volume={39},
  number={24},
  pages={25886--25894},
  year={2025}
}

@article{yang2024exploring,
  title={Exploring and unleashing the power of large language models in automated code translation},
  author={Yang, Zhen and Liu, Fang and Yu, Zhongxing and Keung, Jacky Wai and Li, Jia and Liu, Shuo and Hong, Yifan and Ma, Xiaoxue and Jin, Zhi and Li, Ge},
  journal={Proceedings of the ACM on Software Engineering},
  volume={1},
  number={FSE},
  pages={1585--1608},
  year={2024},
  publisher={ACM New York, NY, USA}
}

@inproceedings{ye2025uncovering,
  title={Uncovering llm-generated code: A zero-shot synthetic code detector via code rewriting},
  author={Ye, Tong and Du, Yangkai and Ma, Tengfei and Wu, Lingfei and Zhang, Xuhong and Ji, Shouling and Wang, Wenhai},
  booktitle={Proceedings of the AAAI Conference on Artificial Intelligence},
  volume={39},
  number={1},
  pages={968--976},
  year={2025}
}

@inproceedings{gao2024search,
  title={Search-based llms for code optimization},
  author={Gao, Shuzheng and Gao, Cuiyun and Gu, Wenchao and Lyu, Michael},
  booktitle={2025 IEEE/ACM 47th International Conference on Software Engineering (ICSE)},
  pages={254--266},
  year={2024},
  organization={IEEE Computer Society}
}

@online{claude4,
  author={Anthropic},
  title={Welcome to Claude 4: Your Partner in AI Innovation},
  url={https://claude4.org/},
  year={2025}
}

@article{zhang2024codedpo,
  title={Codedpo: Aligning code models with self generated and verified source code},
  author={Zhang, Kechi and Li, Ge and Dong, Yihong and Xu, Jingjing and Zhang, Jun and Su, Jing and Liu, Yongfei and Jin, Zhi},
  journal={arXiv preprint arXiv:2410.05605},
  year={2024}
}

@article{paul2025obscuracoder,
  title={ObscuraCoder: Powering Efficient Code LM Pre-Training Via Obfuscation Grounding},
  author={Paul, Indraneil and Yang, Haoyi and Glava{\v{s}}, Goran and Kersting, Kristian and Gurevych, Iryna},
  journal={arXiv preprint arXiv:2504.00019},
  year={2025}
}

@article{huang2024soap,
  title={SOAP: enhancing efficiency of generated code via self-optimization},
  author={Huang, Dong and Dai, Jianbo and Weng, Han and Wu, Puzhen and Qing, Yuhao and Zhang, Jie M and Cui, Heming and Guo, Zhijiang},
  journal={arXiv e-prints},
  pages={arXiv--2405},
  year={2024}
}

@article{wong2024aligning,
  title={Aligning crowd-sourced human feedback for reinforcement learning on code generation by large language models},
  author={Wong, Man Fai and Tan, Chee Wei},
  journal={IEEE Transactions on Big Data},
  year={2024},
  publisher={IEEE}
}

@article{zhang2024plum,
  title={Plum: Improving code lms with execution-guided on-policy preference learning driven by synthetic test cases},
  author={Zhang, Dylan and Diao, Shizhe and Zou, Xueyan and Peng, Hao},
  journal={arXiv preprint arXiv:2406.06887},
  year={2024}
}

@article{du2025afterburner,
  title={Afterburner: Reinforcement Learning Facilitates Self-Improving Code Efficiency Optimization},
  author={Du, Mingzhe and Tuan, Luu Anh and Liu, Yue and Qing, Yuhao and Huang, Dong and He, Xinyi and Liu, Qian and Ma, Zejun and Ng, See-kiong},
  journal={arXiv preprint arXiv:2505.23387},
  year={2025}
}

@article{gee2024code,
  title={Code-optimise: Self-generated preference data for correctness and efficiency},
  author={Gee, Leonidas and Gritta, Milan and Lampouras, Gerasimos and Iacobacci, Ignacio},
  journal={arXiv preprint arXiv:2406.12502},
  year={2024}
}

@article{peng2025coffe,
  title={Coffe: A code efficiency benchmark for code generation},
  author={Peng, Yun and Wan, Jun and Li, Yichen and Ren, Xiaoxue},
  journal={Proceedings of the ACM on Software Engineering},
  volume={2},
  number={FSE},
  pages={242--265},
  year={2025},
  publisher={ACM New York, NY, USA}
}

@inproceedings{coignion2024performance,
  title={A performance study of llm-generated code on leetcode},
  author={Coignion, Tristan and Quinton, Cl{\'e}ment and Rouvoy, Romain},
  booktitle={Proceedings of the 28th international conference on evaluation and assessment in software engineering},
  pages={79--89},
  year={2024}
}

@article{liu2024evaluating,
  title={Evaluating language models for efficient code generation},
  author={Liu, Jiawei and Xie, Songrun and Wang, Junhao and Wei, Yuxiang and Ding, Yifeng and Zhang, Lingming},
  journal={arXiv preprint arXiv:2408.06450},
  year={2024}
}

@article{achiam2023gpt,
  title={Gpt-4 technical report},
  author={Achiam, Josh and Adler, Steven and Agarwal, Sandhini and Ahmad, Lama and Akkaya, Ilge and Aleman, Florencia Leoni and Almeida, Diogo and Altenschmidt, Janko and Altman, Sam and Anadkat, Shyamal and others},
  journal={arXiv preprint arXiv:2303.08774},
  year={2023}
}

\end{document}